\documentclass{rmf-d}
\usepackage{nopageno,rmfbib,multicol,times,epsf,amsmath,amssymb,cite}
\usepackage[latin1]{inputenc}
\usepackage[]{caption2}
\usepackage{graphics}
\usepackage{graphicx}
\usepackage[hidelinks]{hyperref}
\hypersetup{colorlinks, citecolor=black, linkcolor=black ,urlcolor={blue}}
\usepackage{wrapfig, blindtext}
\usepackage{comment}
\usepackage{tabularx}
\usepackage{float}
\usepackage{dblfnote}
\newcommand{\Z}{\mathbb{Z}}

\newcommand{\R}{\mathbb{R}}
\newcommand{\C}{{\kern+.25em\sf{C}\kern-.45em\sf{{\small{I}}} \kern+.45em\kern-.25em}}

\newcommand{\be}{\begin{equation}}
\newcommand{\ee}{\end{equation}}
\newcommand{\bea}{\begin{eqnarray}}
\newcommand{\eea}{\end{eqnarray}}
\newcommand{\nn}{\nonumber}

\def\rmfcornisa{ \hfill\rmf\ {} \hfill }

\def\rmfcintilla{{\it Rev.\ Mex.\ Fis.\/} }
\clearpage \rmfcaptionstyle 
\begin{document}
\title{The structure of cosmic strings of a U(1) gauge field for the
conservation of $B - L$
\vspace{-6pt}}
\author{Victor Mu\~{n}oz-Vitelly}
\author{Jos\'{e} Antonio Garc\'{\i}a-Hern\'{a}ndez}
\author{Wolfgang Bietenholz}
\address{Instituto de Ciencias Nucleares, Universidad Nacional Aut\'{o}noma
de M\'{e}xico \\ A.P.\ 70-543, C.P.\ 04510 Ciudad de M\'{e}xico, M\'{e}xico}
\maketitle
\begin{abstract}
\vspace{1em}
We consider an extension of the Standard Model, where the difference
between the baryon number $B$ and the lepton number $L$ is gauged
with an Abelian gauge field, in order to explain the exact conservation
of $B-L$. To avoid a gauge anomaly, we add a right-handed
neutrino $\nu_{\rm R}$ to each fermion generation. Here it is not
sterile, so the usual Majorana term is excluded by gauge invariance.
We provide a mass term for $\nu_{\rm R}$ by adding a non-standard 1-component
Higgs field, thus arriving at a consistent extension of the Standard
Model, where the conservation of $B-L$ is natural, with a modest
number of additional fields.
We study the possible formation of cosmic strings
by solving the coupled field equations of the two Higgs fields and the
non-standard U(1) gauge field. Numerical methods provide the corresponding
string profiles, depending on the Higgs winding numbers, such that
the appropriate boundary conditions in the string center and far from
it are fulfilled.

\vspace{1em}
\end{abstract}
\keys{cosmic strings, baryon and lepton number, non-standard Higgs field \vspace{-4pt}}
\pacs{11.27.+d, 11.30.Fs, 12.60.-i, 14.70.Pw, 14.80.Fd \vspace{-4pt}}
\begin{multicols}{2}

\section{A modest extension of the Standard Model}

The Standard Model of particle physics is a major scientific achievement
of the 20th century, and of all times --- many of its predictions have
been confirmed to an enormous precision. Still there are some reservations
about it, which often refer to the missing inclusion of gravity and
Dark Matter. Here we take a different point of departure to motivate
a possible extension beyond the Standard Model.

Since the 20th century, symmetries are a central concept of physics.
They can be divided into global and local symmetries. We consider the
latter equivalent to gauge symmetries, which must be exact based on
the foundation of gauge invariance. This property strongly constrains
the options of consistent extensions beyond the Standard Model, because
one has to assure that the gauge anomalies still cancel.

On the other hand, there is no compelling reason for global symmetries 
to be exact. Indeed, they are usually just approximately valid in
some energy regime, where symmetry breaking terms are hardly
manifest, although they exist at a higher energy scale.
An exception is Lorentz invariance, which --- along with locality ---
also implies CPT invariance, but if we invoke gravity (as described
by General Relativity), it turns into a local symmetry, which is
naturally exact.

Another exception, which does not have such a plausibility argument,
is the difference between the baryon number $B$ and the lepton
number $L$. Experimentally no violation of $B$ or of $L$ has
ever been observed, but the Standard Model allows for transitions, which
turn quarks into leptons or vice versa. They are based on topological
windings of the Yang-Mills gauge field SU(2)$_{\rm L}$, which affect
both the left-handed quark- and lepton-doublets. However, this
simultaneous effect still keeps the difference $B-L$ invariant.
This is manifest from the fact that the divergences of the baryon
current $J_{\mu}^{B}$ and the lepton current $J_{\mu}^{L}$ coincide,
\be  \label{JBJLdiv}
\partial^{\mu} J_{\mu}^{B} = \partial^{\mu} J_{\mu}^{L} =
- \frac{N_{\rm g}}{32 \pi^{2}} {\rm Tr} \, [W^{\mu \nu} \tilde W_{\mu \nu}] \ ,
\ee
where $W^{\mu \nu}$ is the SU(2)$_{\rm L}$ field strength tensor,
$\tilde W_{\mu \nu} = \epsilon_{\mu \nu \rho \sigma} W^{\rho \sigma}$ and
$N_{\rm g}$ is the number of fermion generations.

$B-L$ invariance is not a paradox, but it appears strange
that this global symmetry should be ``accidentally'' exact.
This is the conceptually unsatisfactory point that we try to overcome
by going a step beyond the Standard Model. However, we do so in an
economic way, by essentially introducing just the minimum of additional
ingredients which are necessary to render consistency and some natural
features.

Our economic approach proceeds in three steps, which we first
describe in words.

\begin{itemize}

\item We promote the $B-L$ invariance to a gauge symmetry,
  which corresponds to an Abelian Lie group U(1)$_{Y'}$, and we
  denote its gauge field as ${\cal A}_{\mu}$.
  Thus the gauge group structure of the Standard Model is extended
  to ${\rm SU}(3)_{c} \otimes {\rm SU}(2)_{\rm L} \otimes {\rm U}(1)_{Y}
  \otimes {\rm U}(1)_{Y'}$, with dimension 13 and rank 5.  
  ${\cal A}_{\mu}$ could couple just to $B-L$, or
  to a linear combination of the charges $Y$ and $B-L$,
  which are both conserved. We write it as
  \be  \label{Yprime}
  Y' = 2 h Y + \frac{h'}{2} (B-L)
  \ee
  where $h$ and $h'$ are coupling constants
  (the factors of 2 and 1/2 will be convenient later).
  It will become massive (see below), which leads to a heavy
  $Z'$-boson, while ${\rm SU}(2)_{\rm L} \otimes {\rm U}(1)_{Y}$ gives
  rise to the standard gauge bosons $W^{\pm}$, $Z$ and $\gamma$,
  as usual.

\item With this additional field, and the quarks and leptons of
  the (traditional) Standard Model, a gauge anomaly emerges.
  This can easily be seen from a fermionic triangular
  diagram, with a gauge couplings to the charge $B-L$ at each
  vertex, as illustrated in Fig.\ \ref{tria}.
  In each fermion generation, we have 2 quark flavors,
  with a left- and right-handed quark and 3 colors, and
  baryon number $B=1/3$, which sums up to $B=4$.
  In the lepton sector we have one lepton with both chiralities,
  but only a left-handed neutrino, all with lepton number $L=1$,
  which amounts to $L=3$. Cancellation can be achieved by adding
  another lepton to each generation, and the obvious scenario is
  the inclusion of a right-handed neutrino $\nu_{\rm R}$.
\begin{figure}[H]
\centering
\includegraphics[scale=0.1]{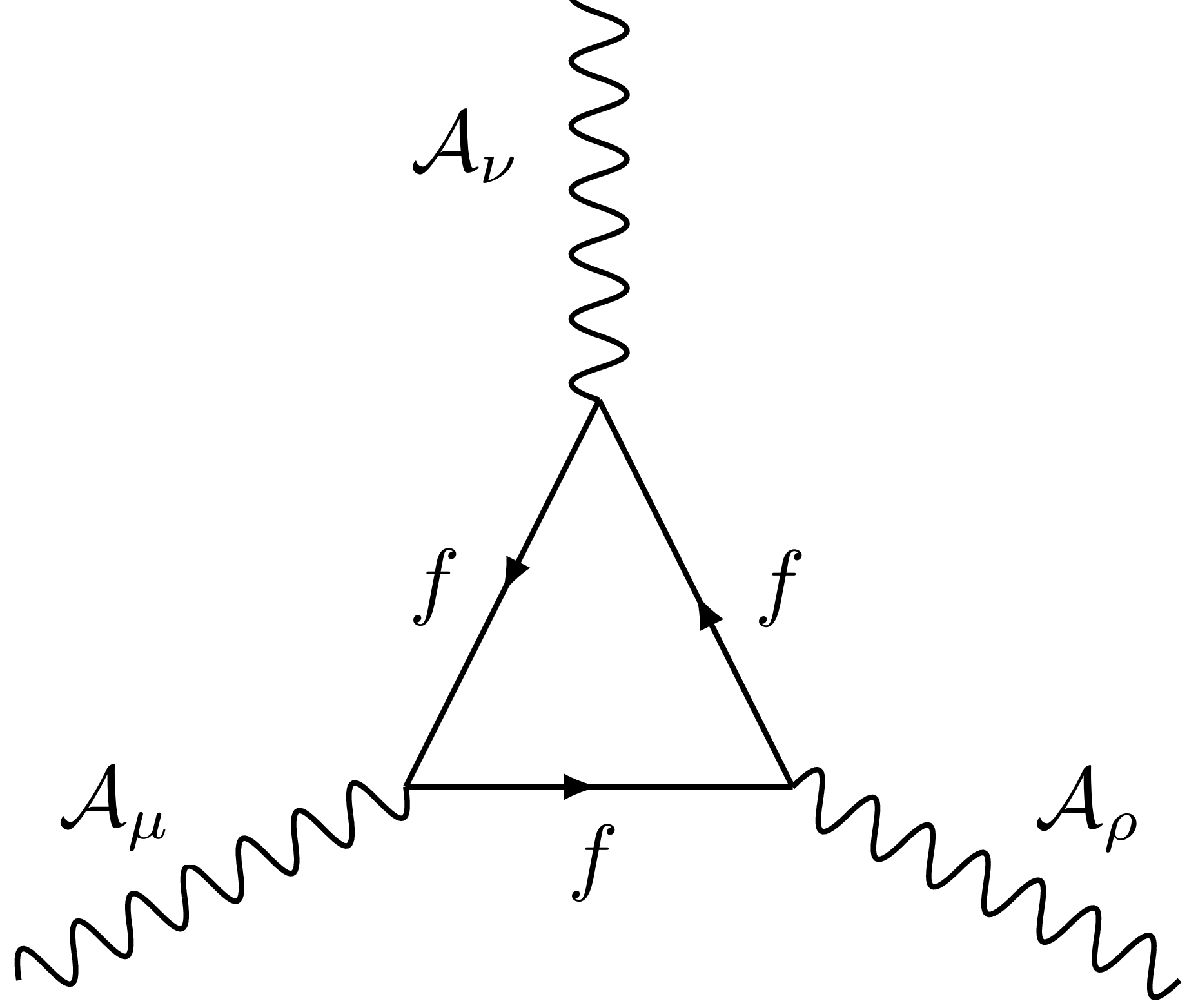}
\caption{A fermionic triangle diagram, where $f$ runs over all fermions
  involved. Their couplings to the external legs ${\cal A}_{\mu}$,
  ${\cal A}_{\nu}$, ${\cal A}_{\rho}$ depend on the quantum number $B-L$.}
\label{tria}
\end{figure}

  It is well-known that this neutrino is ``sterile'' in the sense
  that it does not couple to gauge fields of the Standard Model.
  So it does not affect the anomaly cancellations with respect to
  the Standard Model gauge fields.
  It is also welcome in other respects: it provides the possibility
  to include a neutrino mass, while maintaining renormalizability,
  and it is even a candidate for Dark Matter.

\item With $\nu_{\rm R}$ included in each generation, we can
  build Dirac mass terms for the neutrinos, with the same structure
  as for the quark flavors $u$, $c$ and $t$,
  by a Yukawa coupling of $\nu_{\rm L}$ and $\nu_{\rm R}$ to the
  standard Higgs doublet field $\Phi$. 

  Once $\nu_{\rm R}$ is present, it is natural for it to have also
  a Majorana-type mass term, which is independent of $\nu_{\rm L}$.
  However, in this scenario the usual Majorana term cannot be
  added to the Lagrangian: it is built solely from two factors of
  $\nu_{\rm R}$, hence it has $L=2$. In other scenarios this is
  allowed, but in our case this term is not U(1)$_{Y'}$ gauge invariant.

  In order to be able to construct a Majorana-type (purely right-handed)
  neutrino mass term, we still add a non-standard Higgs field $\chi$.
  In the framework of our economic approach, we assume a 1-component
  complex scalar field, $\chi \in \C$. Now we can add a Majorana-Yukawa
  term $\propto \chi \nu_{\rm R}^{T} \nu_{\rm R} + {\rm c.c.}$ in
  each generation, and gauge invariance holds if $\chi$ carries the
  quantum number $B-L = 2$ ($B$ and $L$ do not need to be specified
  separately).

  Like the standard Higgs field $\Phi \in \C^{2}$, $\chi$ can be
  applied to give mass to $\nu_{R}$ in all fermion generations,
  and it has a quartic (renormalizable) potential.
  This potential gives rise to spontaneous symmetry breaking, with
  a large vacuum expectation value (VEV) $v'$, which arranges for a
  heavy $Z'$-boson.

  Finally, is also natural to include a mixed Higgs term
  $\propto \Phi^{\dagger} \Phi \chi^{*} \chi$.
  
\end{itemize}

So we have designed a modest extension of the Standard Model, with one
additional Abelian gauge field ${\cal A}_{\mu}$,
a right-handed neutrino $\nu_{R}$ in each fermion
generation, plus a 1-component non-standard Higgs field $\chi$.
Each of these fields is hypothetical, but they have a clear motivation,
as we pointed out above.

If we want to embed this model into a Grand Unified Theory (GUT), we
cannot use the unified gauge group SU(5), which only has rank 4, so the
simplest and obvious choice is SO(10) \cite{Mink74}. In this framework,
the non-standard hypercharge $Y'$ takes the specific form
\cite{Buch91}
\be  \label{YprimeSO10}
Y' = Y - \frac{5}{4} (B-L) \ ,
\ee
which we will consider below. This form fulfills the orthonormality
conditions $\sum_{f} Y_{f} Y'_{f} = 0$, $\sum_{f} Y^{2}_{f}
= \tfrac{2}{3} \sum_{f} Y'^{\, 2}_{f} = \tfrac{10}{3} N_{\rm g}$,
where the sums run over all fermions \cite{Buch91}.

A fully-fledged quantum field analysis of this model
is beyond the scope of this work.
We are going to discuss numerical solutions to the coupled field
equations of the two Higgs fields and the U(1)$_{Y'}$ gauge field
(without including fermions and standard gauge fields; fermionic
contributions to cosmic strings are discussed {\it e.g.}\
in Ref. \cite{WQG}).
In this semi-classical analysis, we are interested in the
possible formation of cosmic strings due to topological defects,
where the two Higgs fields may have arbitrary (integer) winding
numbers. Before explicitly addressing the corresponding field
equations, we insert some general remarks about cosmic strings.

\section{Topological defects and cosmic strings}

Topological defects are relevant in a variety of condensed matter
systems, as reviewed {\it e.g.}\ in Ref.\ \cite{Victor}. They appear
for instance in type II superconductors \cite{supcon2}, and in some
cases their percolation is related to a phase transition.

The possibility of the mathematically analogous formation of
cosmic strings was first considered by Kibble in 1976 \cite{Kibble}.
This scenario has attracted attention ever since, although there is
no evidence so far for the physical existence of cosmic strings
(bounds on the string tension are obtained in particular from
the Cosmic Microwave Background \cite{CMB}). The recent search
for evidence of cosmic strings focuses on the detection of
gravitational waves \cite{LIGO}.

Regarding the early Universe, there are attempts to relate cosmic
strings with the electroweak phase transition
(the electroweak symmetry remains unbroken in the core) \cite{Brand}.
For a stability discussion of electroweak strings, see {\it e.g.}\
Ref.\ \cite{James}.

Cosmic strings could also be superconducting \cite{supcos},
which would establish a direct link to condensed matter physics.

\begin{figure}[H]
\vspace{-2mm}
\centering
\includegraphics[scale=0.35]{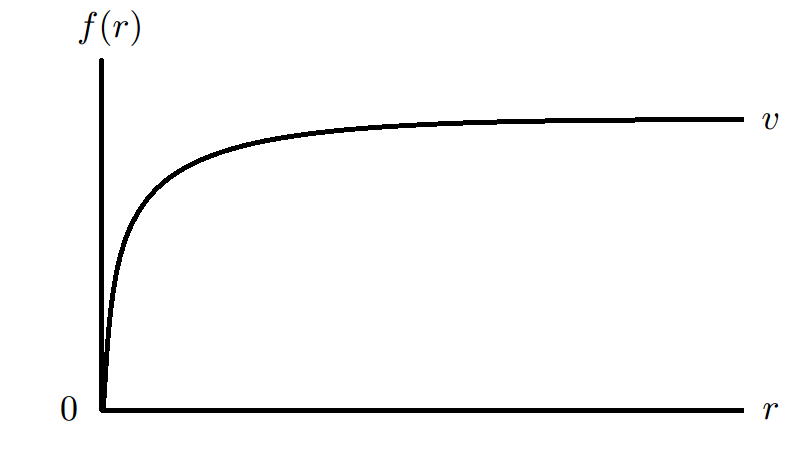}
\caption{A prototype profile function $f(r)$ of a cosmic string, with
  a topological defect at $r=0$, and $f(r \to \infty) = v$
  (in cylindrical coordinates).}
\label{profile}
\end{figure}

Here we briefly sketch the basic idea; for extensive reviews we refer
to Refs.\ \cite{HindKib,Vilenkin}.
As a toy model, we consider a 1-component complex scalar field
$\phi (x)$, with the well-known Lagrangian
\be
   {\cal L}(\phi, \partial_{\mu} \phi) = \partial_{\mu} \phi \partial^{\mu} \phi
   - \mu_{0}^{2} \phi^{*} \phi - \lambda_{0} (\phi^{*} \phi)^{2} \ .
\ee
For $\mu_{0}^{2} < 0$ we obtain a continuous set of classical vacua,
\be
\phi = v \, e^{i \alpha} \ , \quad
v = \sqrt{-\mu_{0}^{2}/2 \lambda_{0}} \ ,
\ee
with an arbitrary phase $\alpha \in (-\pi , \pi]$. It can be generalized
to static solutions with non-minimal energy, for which we write --- in
cylindrical coordinates --- the ansatz
\be
\phi (r,\varphi,z) = f(r) e^{i n \varphi} \ ,
\ee
where $n \in \Z$ is the winding number. For $n \neq 0$ this
represents a topological defect, and the profile function
$f(r)$ should vanish at $r=0$ (this avoids a phase ambiguity).
Far from this defect we require $f(r \to \infty) = v$. These
are the boundary conditions for the field equation
\be
\frac{d^{2}f}{dr^{2}} + \frac{1}{r} \frac{df}{dr} =
\Big( \frac{n^{2}}{r^{2}} + \mu_{0}^{2} + 2 \lambda_{0} \Big) f \ .
\ee
The shape of a typical solution is depicted in Fig.\ \ref{profile}.

\section{Field equations for the extended gauge-Higgs sector}

According to the description in Section 1, we consider the
gauge-Higgs Lagrangian
\bea
   {\cal L} &=& D_{\mu} \Phi^{\dagger} D^{\mu} \Phi +
   d_{\mu} \chi^{*} d^{\mu} \chi \nn \\ && - \frac{1}{4}
   {\cal F}_{\mu \nu} {\cal F}^{\mu \nu} - V(\Phi, \chi) \nn \\
   {\cal F}_{\mu \nu} &=& \partial_{\mu} {\cal A}_{\nu} -
   \partial_{\nu} {\cal A}_{\mu} \nn \\
V(\Phi, \chi) &=& \mu^{2} \Phi^{\dagger} \Phi + \lambda (\Phi^{\dagger} \Phi )^{2}
   +  \mu'^{\, 2} \chi^{*} \chi + \lambda' (\chi^{*} \chi )^{2} \nn \\
   && - \kappa  \Phi^{\dagger} \Phi \chi^{*} \chi \ .
\eea
The potential must be bounded from below, which requires
$\lambda , \, \lambda ' >0$, and $\kappa < 2 \sqrt{\lambda \lambda'}$.
Spontaneous symmetry breaking of both Higgs fields, with VEVs $v,\,v' >0$,
further requires
\be  \label{condSSB}
\mu^{2}, \, \mu'^{\, 2} < 0 \ , \quad
\kappa > {\rm max} \Big(-\frac{2\mu^{2}\lambda'}{\mu'^{\, 2}},
       -\frac{2\mu'^{\, 2}\lambda} {\mu^{2}} \Big) \ . \ 
\ee
Together this implies $\kappa^{2} < 4 \lambda \lambda'$.

$\Phi$ and $\chi$ are the standard and non-standard Higgs fields,
with the charges $Y_{\Phi} = 1/2$, $(B-L)_{\chi} = 2$, hence
(ignoring the SU(2)$_{\rm L}$ and U(1)$_{Y}$ gauge fields)
the covariant derivatives take the form
\be
D_{\mu} \Phi = \partial_{\mu} \Phi + i h {\cal A}_{\mu} \Phi \ ,
\quad
d_{\mu} \chi = \partial_{\mu} \chi + i h' {\cal A}_{\mu} \Phi \ ,
\ee
where we refer to our convention (\ref{Yprime}).

Following the lines of Section 2, we make an ansatz for static solutions
in terms of cylindrical coordinates,
\be
\Phi = \begin{pmatrix} 0 \\ 1 \end{pmatrix} \phi(r) e^{in\varphi} \ ,
\quad \chi = \xi(r) e^{in'\varphi} \ , \quad
\mathcal{A}_\mu =\frac{a(r)}{r} \hat \varphi \ , \nn
\ee
where $\phi,\, \xi,\, a \in \R$ are the radial profile
functions of the field configurations, $n,\, n' \in \Z$ are the winding
numbers of the two Higgs fields, and $\hat \varphi$ is the tangential
unit vector.

In these terms, the Euler-Lagrange field equations read \cite{Victor,JA}
\begin{gather}
  \frac{d^2 \phi}{dr^2} +\frac{1}{r} \frac{d\phi}{dr} =
  \Big[ \frac{(n + h a)^2}{r^2}  + \mu^2 + 2\lambda \phi^2 -
  \kappa \xi^2 \Big] \phi , \notag \\
  \frac{d^2 \xi}{dr^2} + \frac{1}{r} \frac{d\xi}{dr} =
  \Big[ \frac{(n' + h' a)^2}{r^2} + \mu'^{\, 2} + 2 \lambda' \xi^2
   - \kappa \phi^2 \Big] \xi , \notag \\
  \frac{d^2 a}{dr^2}-\frac{1}{r}\frac{da}{dr} =
  2h \phi^2 (n + ha) + 2h' \xi^2 (n' + h'a) \ .
\label{fieldeqs}
\end{gather}

The boundary conditions in the core and asymptotically far from
it --- where the Higgs profile functions become constant --- are
summarized in the following table.
\vspace*{2mm}

\begin{center}
\small{\renewcommand{\arraystretch}{1.3}
\renewcommand{\tabcolsep}{1.35pc}
\begin{tabular}{|c|c|c|}
\hline
 & $r=0$ & $r \to \infty$ \\
\hline
$\phi (r) $ & $0$ \ (if $n \neq 0$) & $v = \sqrt{- \mu^{2}/2 \lambda}$ \\
$\chi (r) $ & $0$ \ (if $n' \neq 0$) & $v' = \sqrt{- \mu'^{\, 2}/2 \lambda'}$\\
$a(r)$    & $0$ & $- n/h = - n'/h'$ \\
          &     & (if $ n\neq 0$ and $n'\neq 0$) \\ 
\hline
\end{tabular}}
\end{center}

\section{Profiles of U(1)$_{Y'}$ cosmic strings}

In the following we are going to show a set of solutions to the
system of coupled second order differential equations given in
eq.\ (\ref{fieldeqs}). They are based on Refs.\ \cite{Victor,JA},
and obtained with the Python function
{\tt scipy.integrate.solve\_bvp}, which applies the damped Newton
method. Consistency tests show that the errors occur mostly close
to $r=0$ (where we have to deal with removable singularities),
but even there they attain at most ${\cal O}(10^{-3})$, and for
$r \geq {\cal O}(1)$ they decrease by several orders of magnitude.
The solutions have also been reproduced with the Relaxation
method and with the Runge-Kutta 4-point method, although the
latter has difficulties in attaining stable plateaux at large $r$.

A solution is uniquely specified when we fix the Higgs self-couplings
$\lambda$ and $\lambda'$, the Higgs field winding numbers $n$ and $n'$
and the asymptotic large-$r$ values of the three fields. We choose
them by fixing directly $v$ and $v'$, as well as the gauge couplings
$h$ and $h'$, with the constraint $n/h = n'/h'$, which is given in
the table of Section 3, and which is obvious from the last line in
eq.\ (\ref{fieldeqs}).

\begin{figure}[H]
\centering
\includegraphics[scale=0.25]{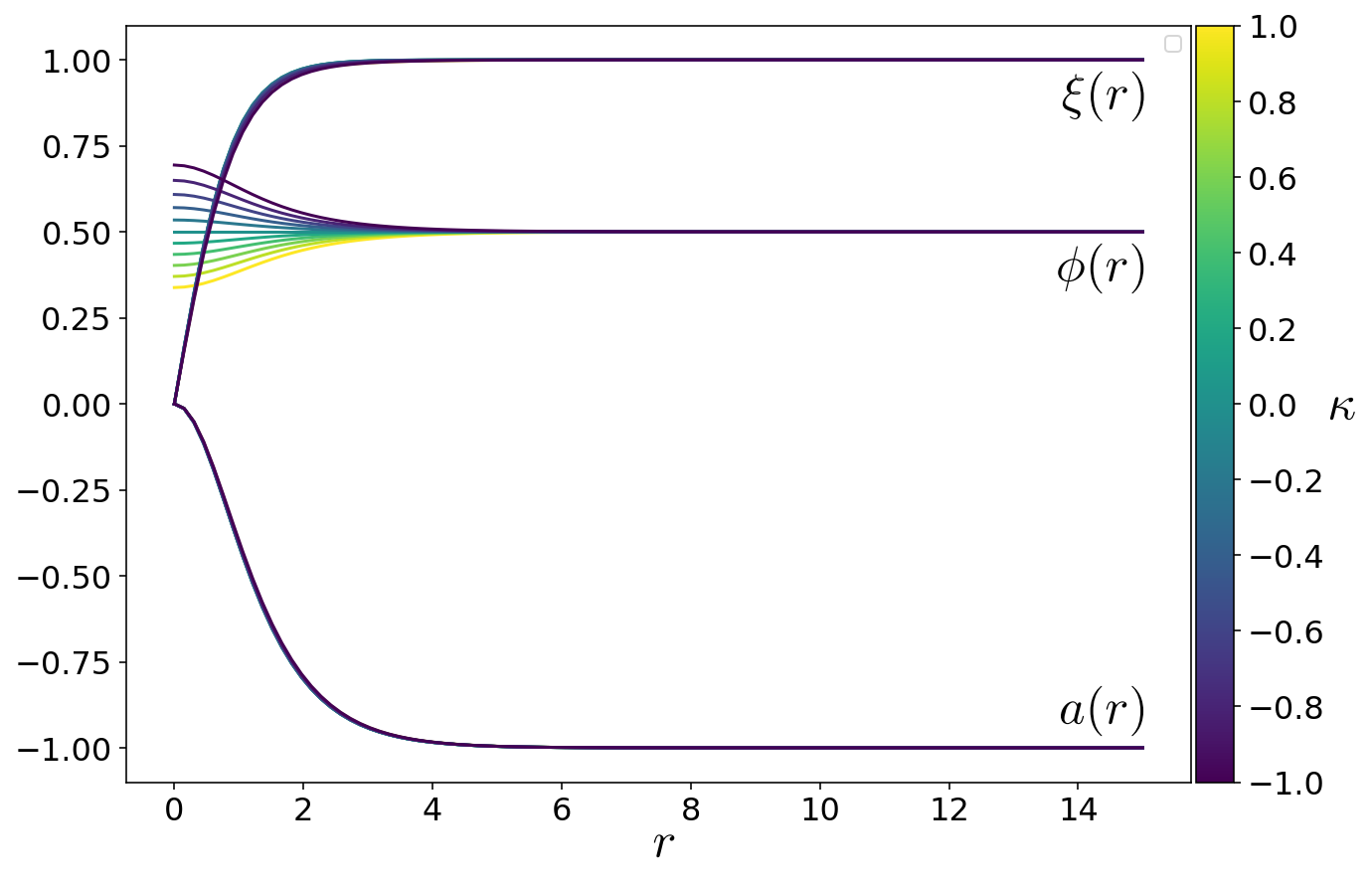}
\caption{Solutions for the cosmic string profile functions,
  with static fields $\Phi$, $\chi$ and ${\cal A}_{\mu}$,
  in cylindrical coordinates, for the parameter set $v=0.5$, $v'=1$;
  $\lambda= \lambda' = 1$; $n = 0$, $n'=1$; $h=0$, $h'=1$.}
\label{fig:CS1}
\end{figure}

\begin{figure}[H]
\centering
\includegraphics[scale=0.25]{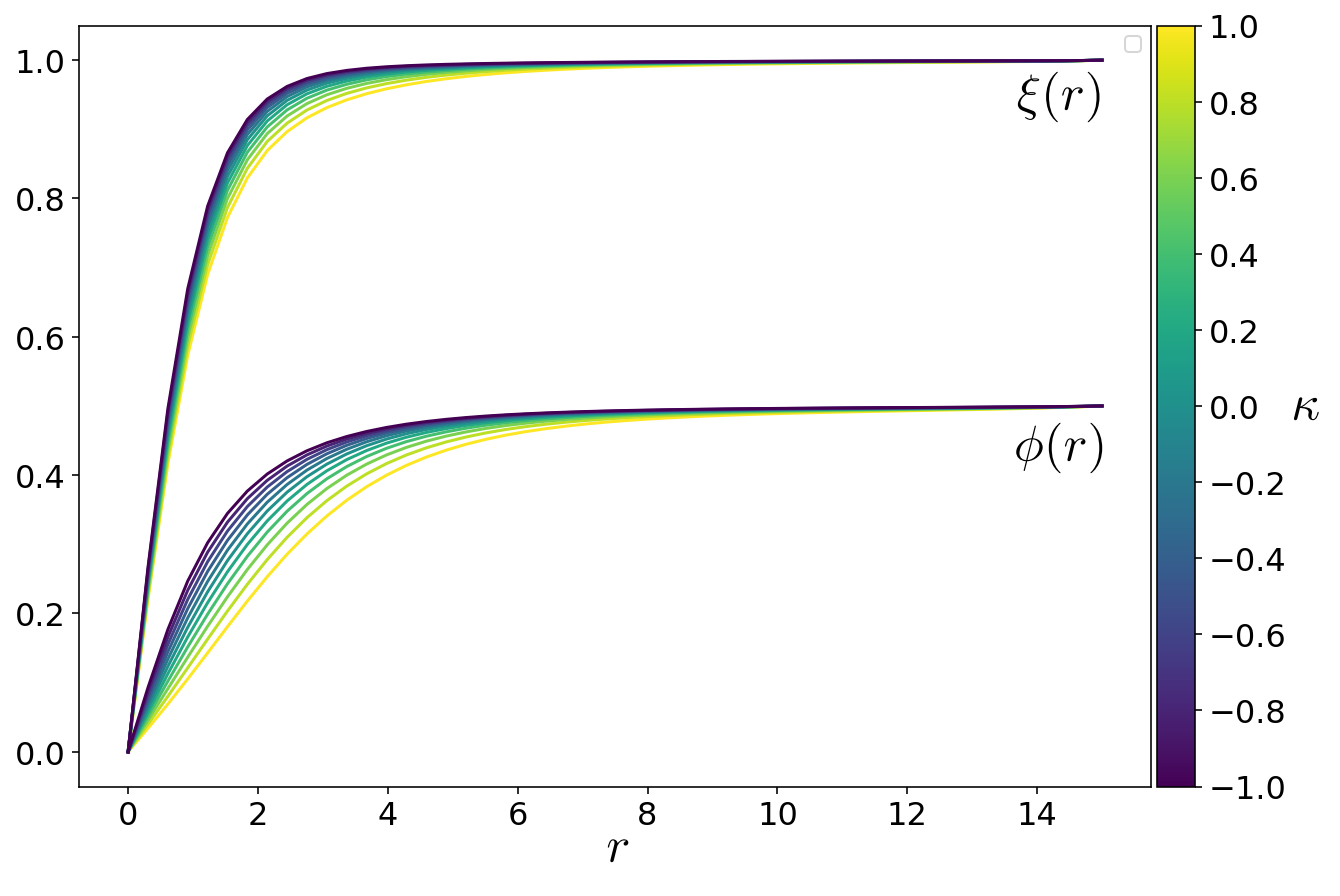}
\includegraphics[scale=0.25]{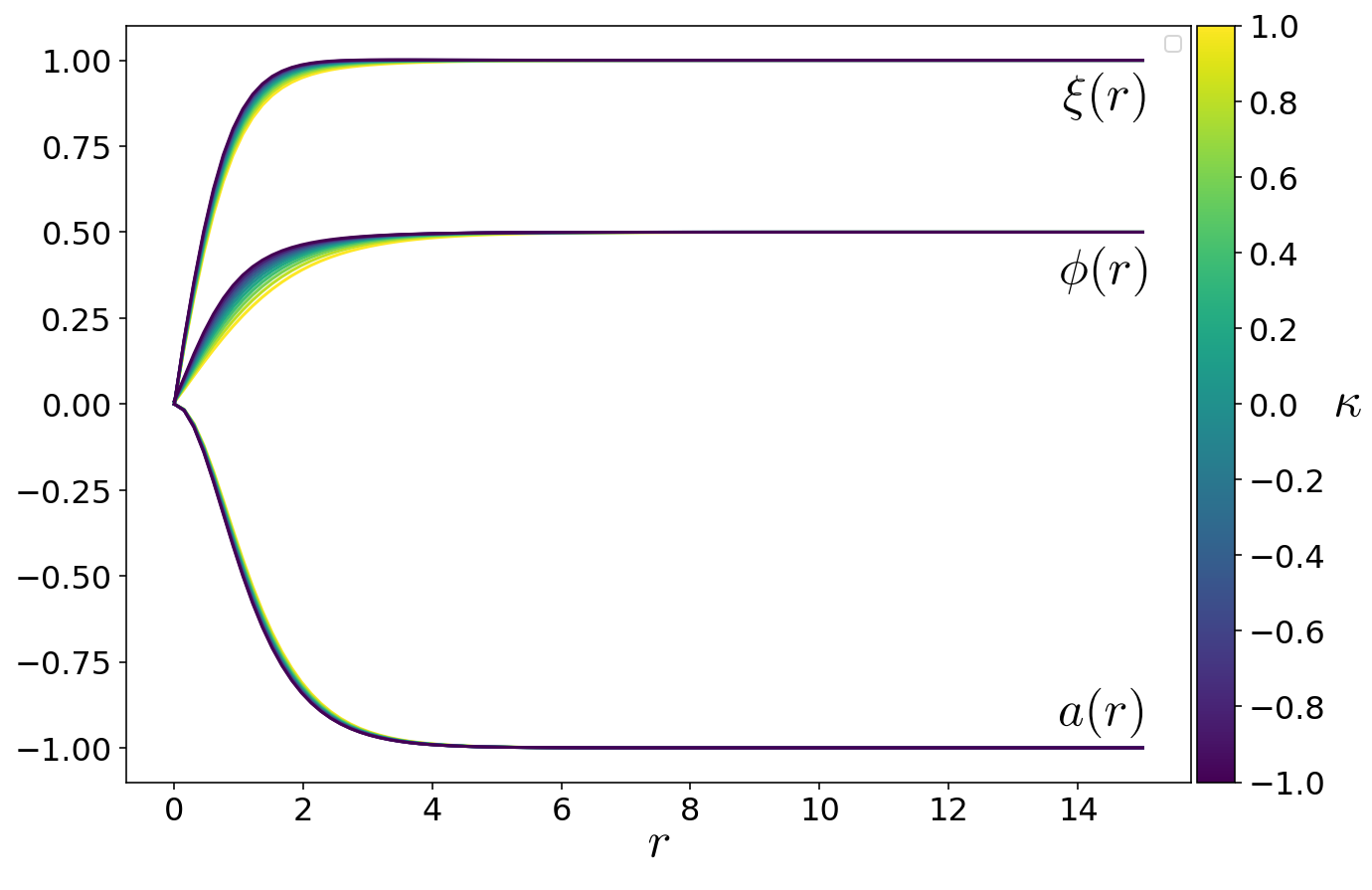}
\caption{Solutions for the cosmic string profile functions,
  with static fields $\Phi$, $\chi$, in the absence (above)
  or presence (below) of a (static) U(1)$_{Y'}$ gauge field ${\cal A}_{\mu}$.
  These plots refer to the parameter set $v=0.5$, $v'=1$;
  $\lambda= \lambda' = 1$; $n = n' = 1$; $h= h'=1$.}
\label{fig:CS2}
\end{figure}

\begin{figure}[H]
\centering
\includegraphics[scale=0.25]{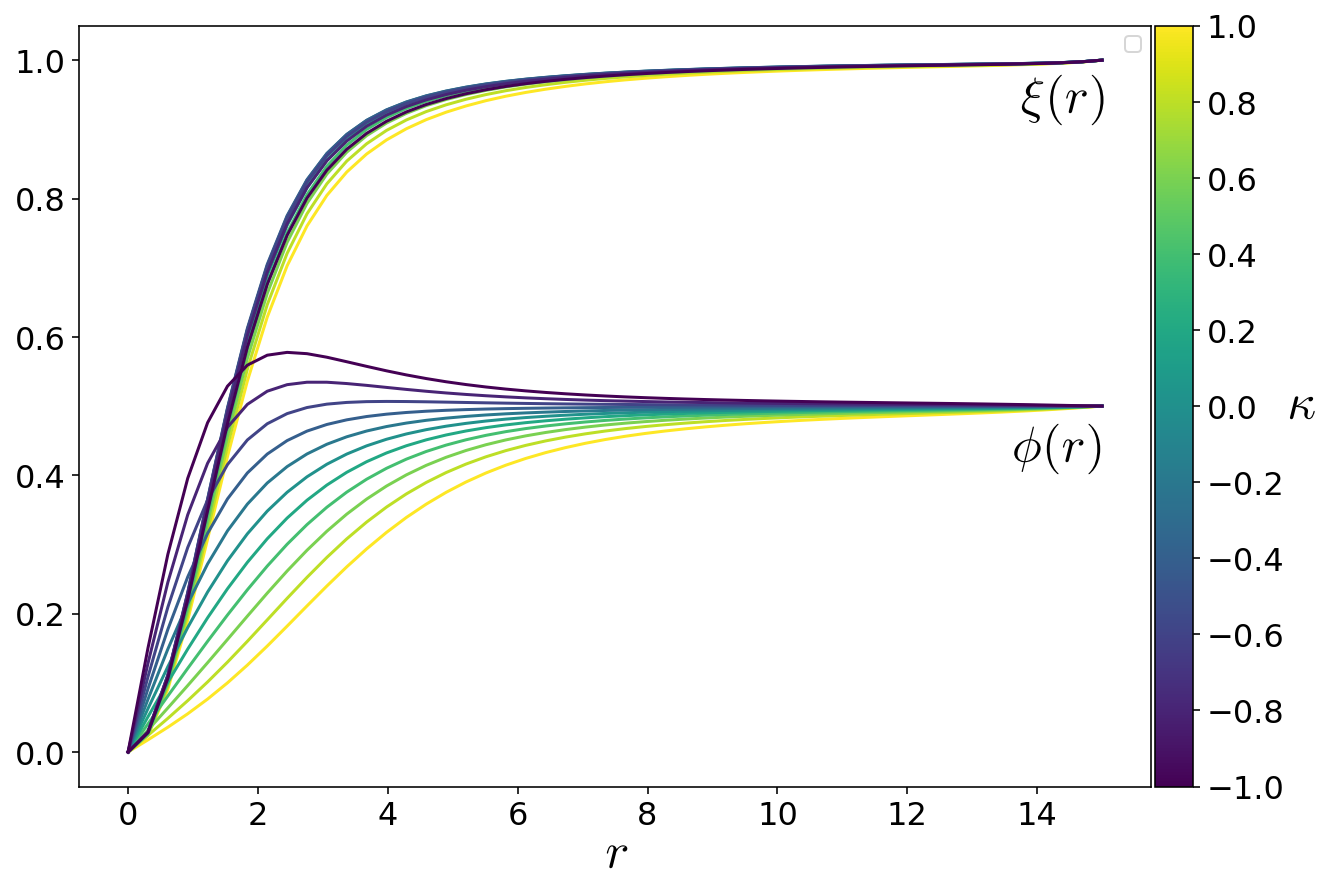}
\includegraphics[scale=0.25]{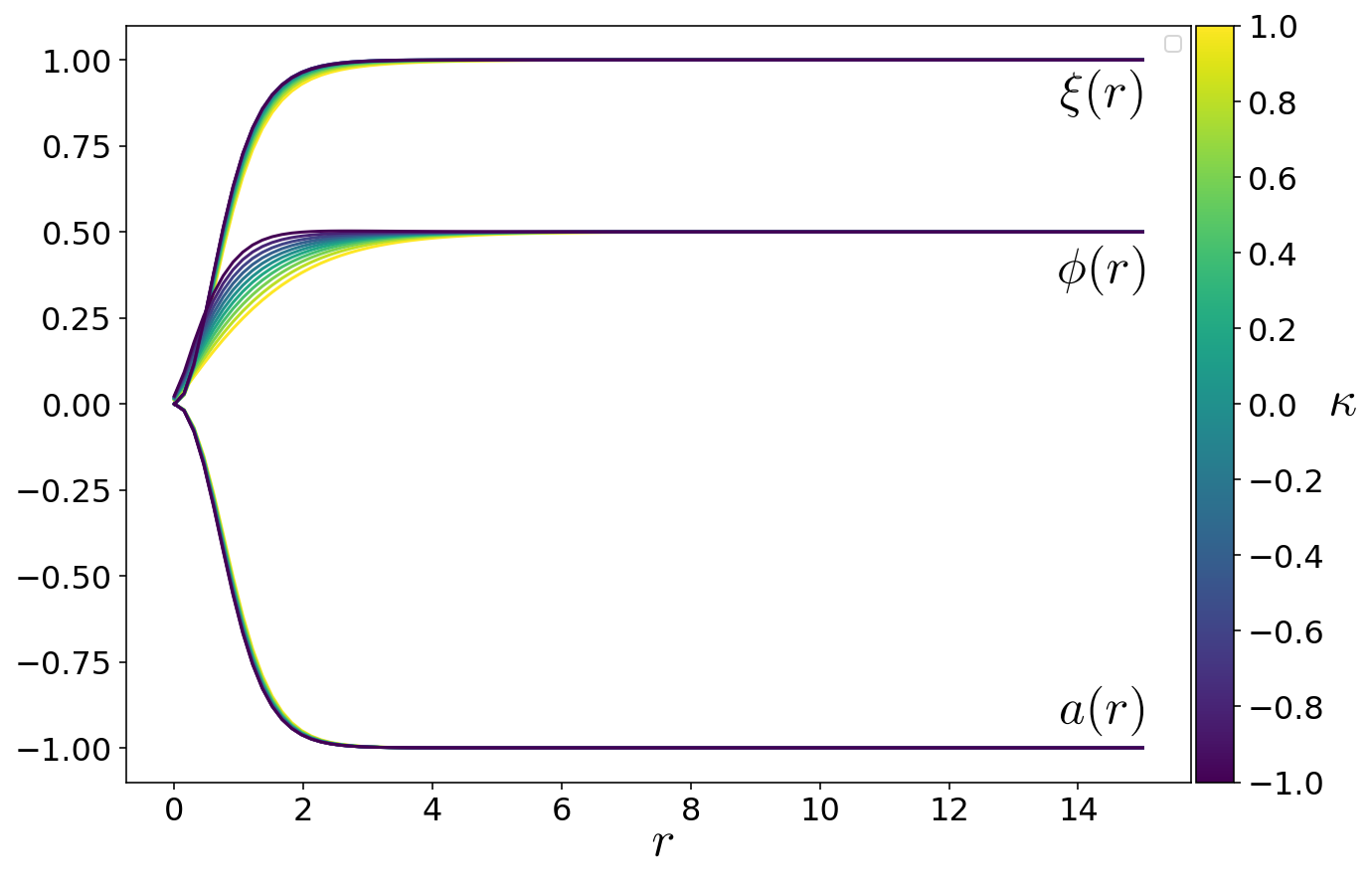}
\caption{Solutions for the cosmic string profile functions,
  with static fields $\Phi$, $\chi$, in the absence (above)
  or presence (below) of a (static) U(1)$_{Y'}$ gauge field ${\cal A}_{\mu}$.
  These plots refer to the parameter set $v=0.5$, $v'=1$;
  $\lambda= \lambda' = 1$; $n = 1$, $n'=2$; $h=1$, $h'=2$.}
\label{fig:CS3}
\end{figure}

In Figs.\ \ref{fig:CS1} to \ref{fig:CS3} we choose the parameters
of the Higgs potentials as
\be  \label{paraset1}
v = 0.5 \ , \ v' = 1 \ , \quad \lambda = \lambda' = 1 \ ,
\quad \kappa \in [-1,1] \ .
\ee
Thus the non-standard Higgs field has a larger VEV
than the one of the Standard Model (in physical units: $v \simeq
246\,{\rm GeV}$), hence the $Z'$-boson tends to be heavy, although this
also depends on its coupling constants $h$ and $h'$. The parameter
$\kappa$ is in the range, which is allowed by the condition for the
potential to be bounded from below, $\kappa < 2$.

If we take the value of $v$ as a reference to convert all quantities
to physical units, then the length unit of $r$ corresponds to
$0.0008\,{\rm fm}$, so our cosmic string solutions have radii of
${\cal O}(10^{-3})\,{\rm fm}$, as the following figures show,
while other theoretical scenarios arrive at larger radii up to
${\cal O}(1)\, {\rm fm}$.

Fig.\ \ref{fig:CS1} shows the case where ${\cal A}_{\mu}$ only couples
to the charge $B-L$, hence $\phi (r=0)$ does not need to vanish.
We see that it deviates only mildly from $\phi (r \to \infty) =v$,
but only for $\phi (r \lesssim 4)$ the value of $\kappa$ is relevant.
Here $a(r \to \infty)$ is fixed by choosing the $\chi$-winding number
$n'=1$, and the coupling $h'=1$.
The profile functions $\xi (r)$ and $a(r)$ move from $0$ to their
large-$r$ values in a manner, which is qualitatively compatible
with the feature of the prototype in Fig.\ \ref{profile}. 

In Fig.~\ref{fig:CS2}
we proceed to the winding numbers $n=n'=1$.
This requires the same couplings, which we choose as $h=h'=1$. Here
we show the cases without or with the presence of the ${\cal A}_{\mu}$
gauge field, {\it i.e.}\ $B-L$ conservation is a global or a local
symmetry, respectively. The behavior of $\xi(r)$ and $\phi(r)$ is
similar in these two cases, but the U(1)$_{Y'}$ gauge symmetry causes
a faster convergence to the large-$r$ limit as $r$ increases,
{\it i.e.}\ it reduces the characteristic string radius,
and it suppresses the $\kappa$-dependence.

In Fig.\ \ref{fig:CS3} we consider a higher winding number of the
$\chi$-field, $n'=2$, again for the cases of a global and local
$B-L$-symmetry. (We recall that the conservation of the weak hypercharge
$Y$ is local in any case due to the gauge group U(1)$_{Y}$, which is
always present, although it is not included in our field equation
analysis.) The impact of the U(1)$_{Y'}$ gauge group is consistent
with Fig.\ \ref{fig:CS2}: it accelerates the convergence of $\xi(r)$
and $\phi(r)$ to their plateau values as $r$ increases, and it
reduces to impact of the $\Phi$-$\chi$-mixing term, {\it i.e.}\ of
the parameter $\kappa$.

In the global symmetry case, the latter is most relevant around
$r\approx 2$, where we observe an interesting phenomenon:
the standard Higgs profile can {\em overshoot,} {\it i.e.}\ exceed
the value of $v$ at a moderate radius $r$.
In addition, we see at small $r \in {\cal O}(0.1)$ that $n'=2$ keeps
$\xi (r)$ small, in particular in the absence of ${\cal A}_{\mu}$,
before it turns to the regime of maximal slope at $r \in {\cal O}(1)$. 
We will come back to this point.

\begin{figure}[H]
\centering
\includegraphics[scale=0.25]{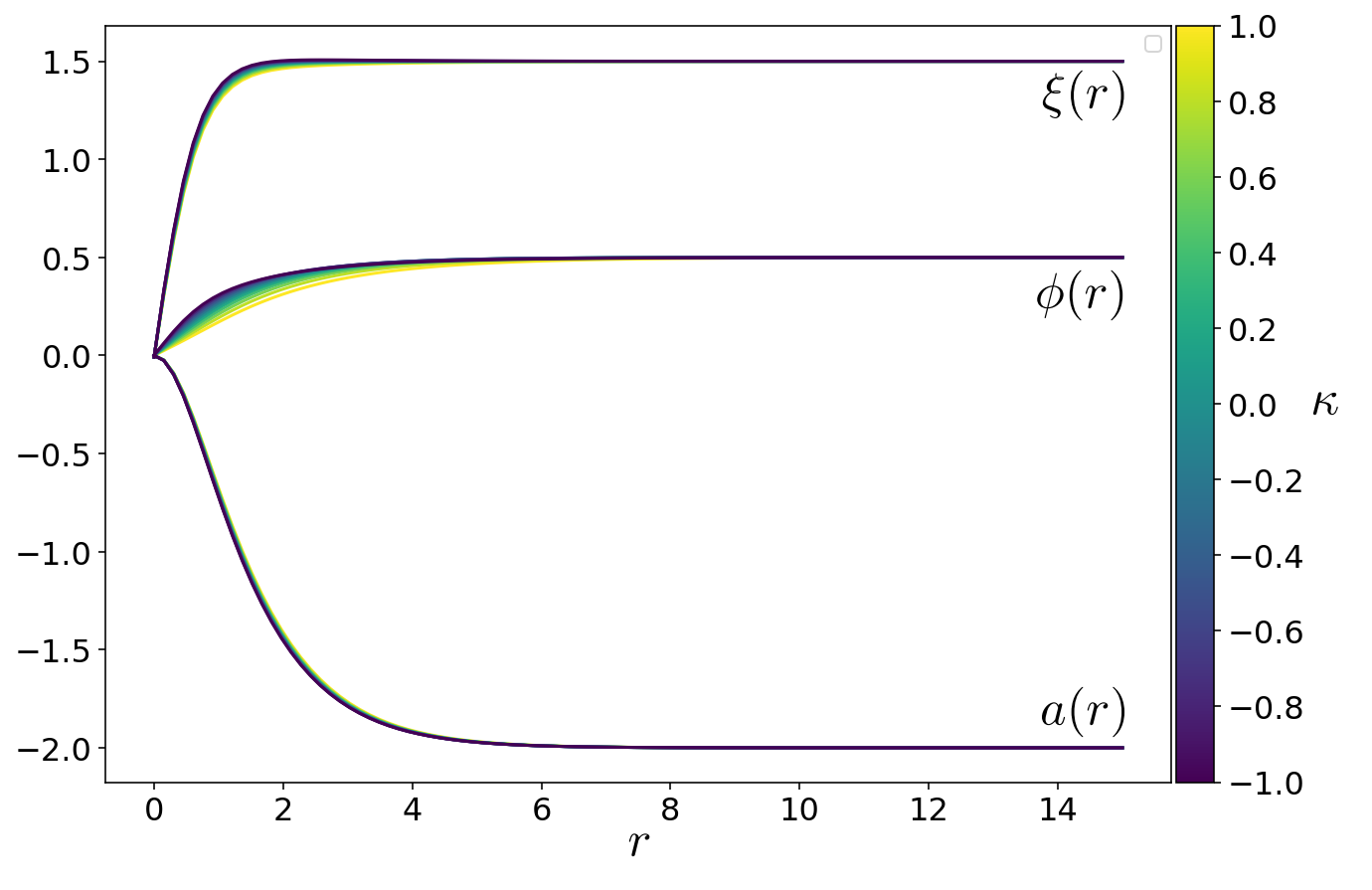}
\caption{Solutions for the cosmic string profile functions,
  with static fields $\Phi$, $\chi$ and ${\cal A}_{\mu}$,
  for the parameter set $v=0.5$, $v'=1.5$;
  $\lambda=0.5$, $\lambda' = 1$; $n=n'=1$; $h=h'=0.5$.}
\label{fig:CS4}
\end{figure}

\begin{figure}[H]
\centering
\includegraphics[scale=0.25]{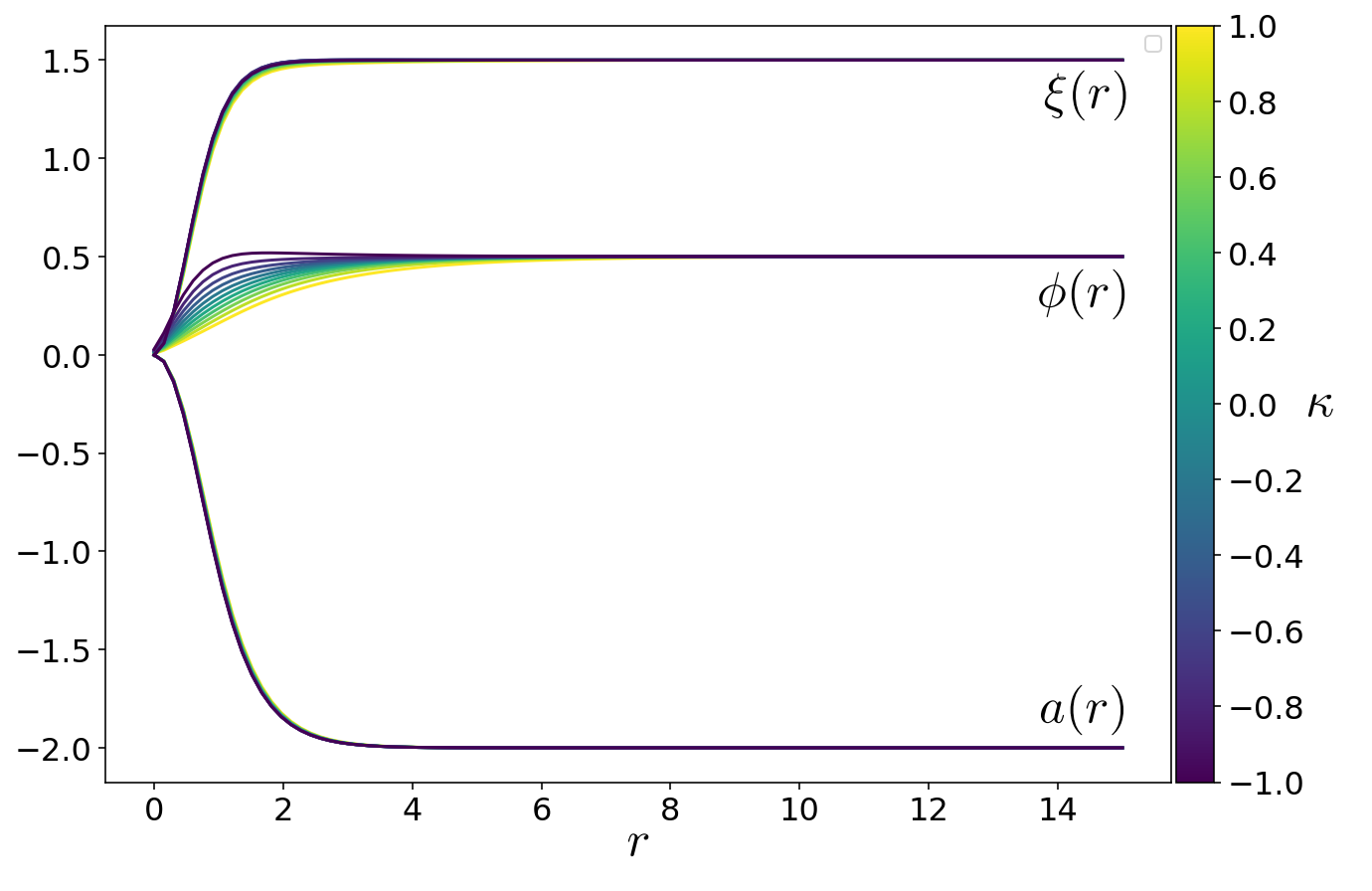}
\caption{Solutions for the cosmic string profile functions,
  with static fields $\Phi$, $\chi$ and ${\cal A}_{\mu}$,
  for the parameter set $v=0.5$, $v'=1.5$;
  $\lambda=0.5$, $\lambda' = 1$; $n=1$, $n'=2$; $h=0.5$, $h'=1$.}
\label{fig:CS5}
\end{figure}

Figs.\ \ref{fig:CS4} and \ref{fig:CS5} go beyond the parameter set
(\ref{paraset1}) by enhancing the ratio $v'/v$ and assuming different
Higgs field self-couplings,
\be
v = 0.5 \ , \ v' = 1.5 \ , \quad \lambda = 0.5 \ , \ \lambda' = 1 \ ,
\quad \kappa \in [-1,1] \ .
\label{paraset2}
\ee
Fig.\ \ref{fig:CS4} refers to the case $n=n'=1$, but we also modify
the couplings compared to Fig.\ \ref{fig:CS2}. The condition is only
that they have to coincide, so we now set them to $h=h'=0.5$.
The qualitative features agree with Figs.\ \ref{fig:CS1} to
\ref{fig:CS3}, in particular in the presence of the gauge field
${\cal A}_{\mu}$, which shows that these features are quite robust
against parameter modifications.

Fig.\ \ref{fig:CS5} returns to the case of the double winding of the
$\chi$-field, with $n=1$, $n'=2$; $h=0.5$, $h'=1$. This plot shows in
particular that the overshooting effect of the profile $\phi (r)$ can
also occur in the presence of the U(1)$_{Y'}$ gauge field, which
strengthens the relevance of this observation.
At $r \in {\cal O}(0.1)$
we observe again a suppression of $\xi(r)$, similar to the lower plot
in Fig.\ \ref{fig:CS3} (which also refers $n'=2$ in the presence of
${\cal A}_{\mu}$). Will will probe this property further at even larger
values of $|n'|$.

Let us finally return to the aforementioned scenario where we consider
this model as a subset of the SO(10) GUT. In this case, the couplings
$h$ and $h'$ are fixed by eq.\ (\ref{YprimeSO10}), which also
implies a specific ratio between the winding numbers,
\be  \label{nprimen}
n' = -5 n \ .
\ee
(There are other conventions in the literature, where $Y'$ differs
by a constant factor, but relation (\ref{nprimen}) does not
depend on it.) This relation also takes us to examples of a large
winding number, which we have not addressed so far.

\begin{figure}[H]
\centering
\includegraphics[scale=0.25]{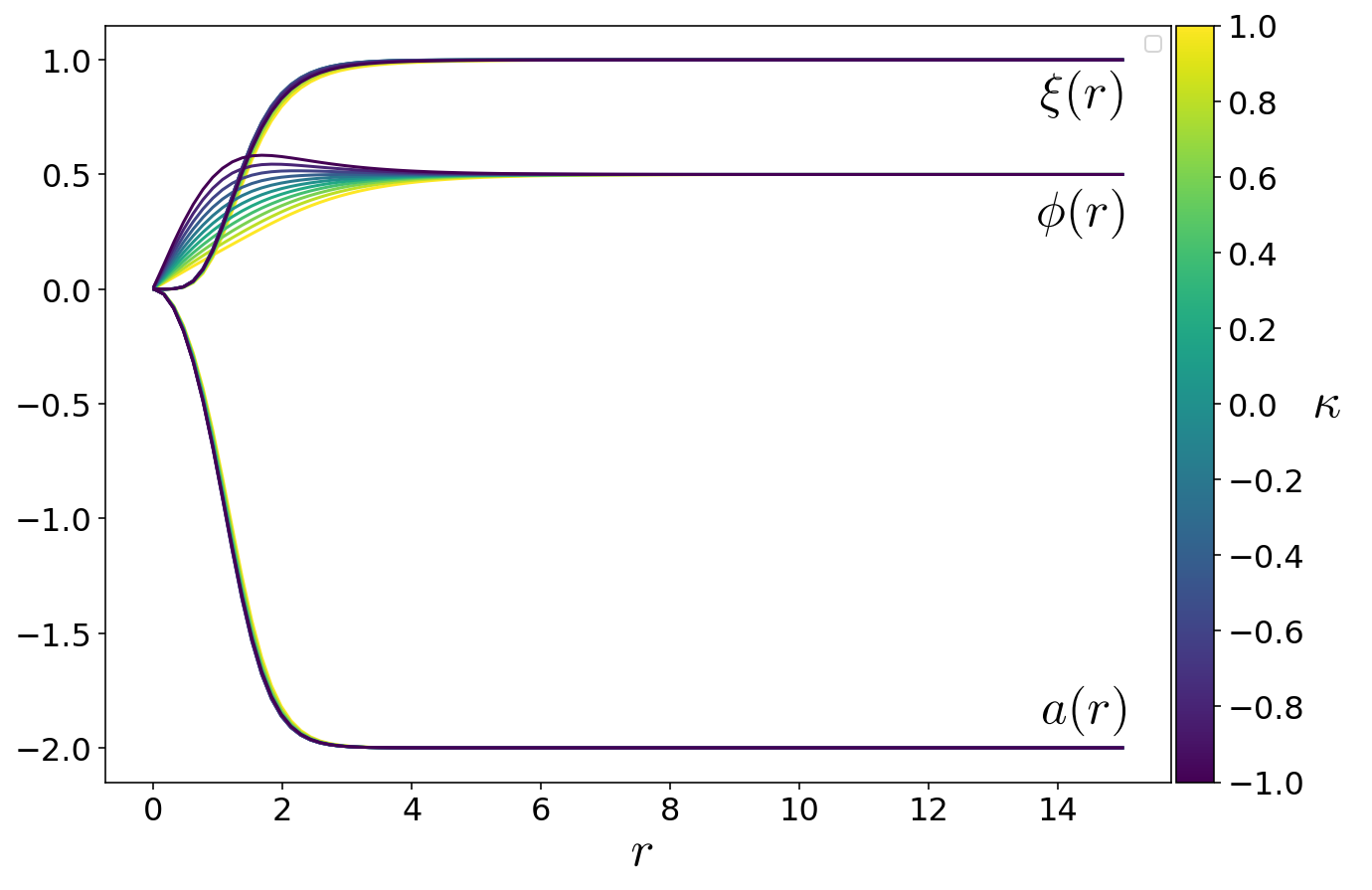}
\caption{Solutions for the cosmic string profile functions,
  for static fields $\Phi$, $\chi$ and ${\cal A}_{\mu}$,
  for the parameter set $v=0.5$, $v'=1$;
  $\lambda=1$, $\lambda' = 1$; $n=1$, $n'=-5$; $h=0.5$, $h'=-2.5$.}
\label{fig:SO10a}
\end{figure}

\begin{figure}[H]
\centering
\includegraphics[scale=0.25]{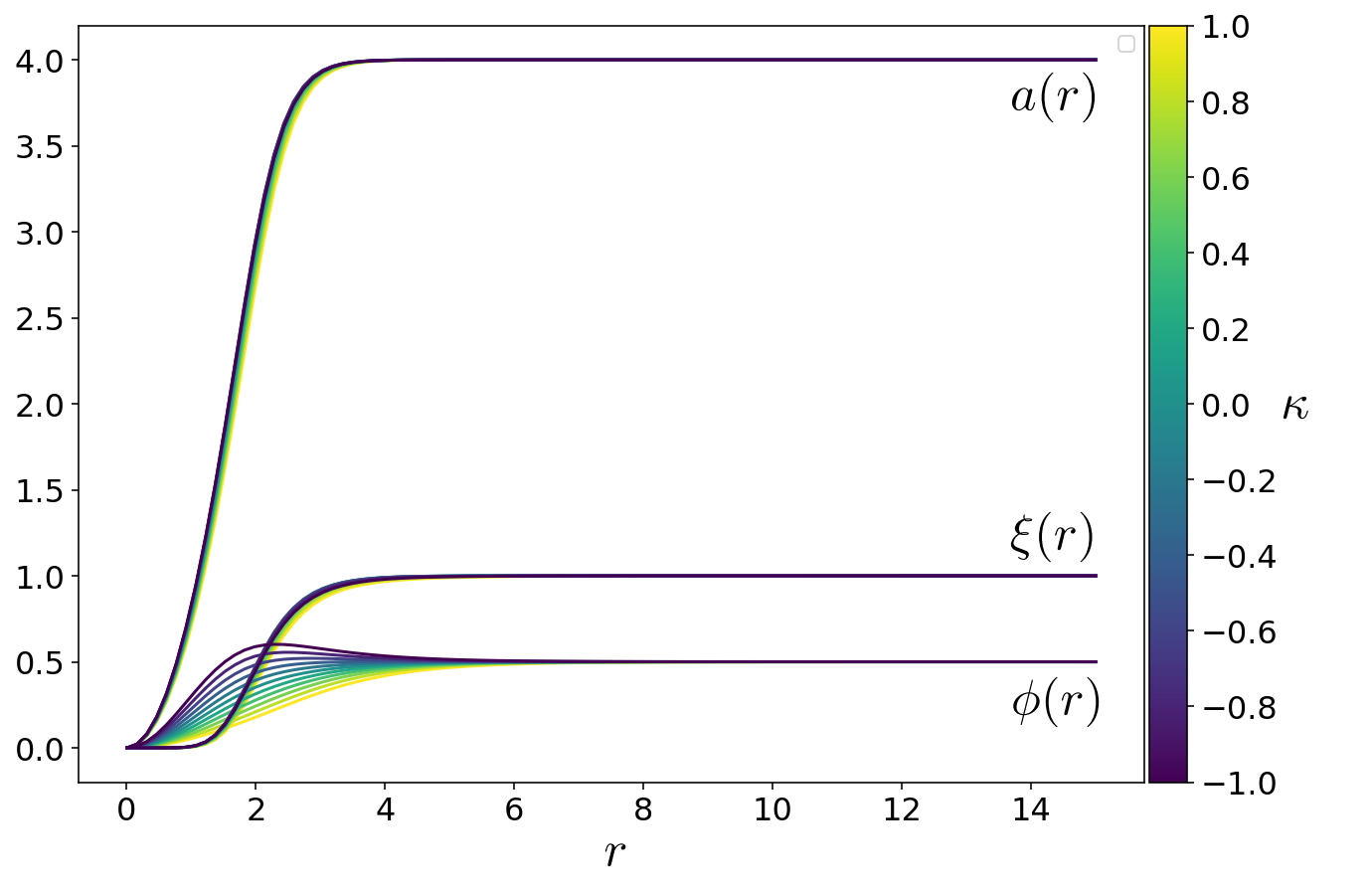}
\caption{Solutions for the cosmic string profile functions,
  for static fields $\Phi$, $\chi$ and ${\cal A}_{\mu}$,
  for the parameter set $v=0.5$, $v'=1$;
  $\lambda=1$, $\lambda' = 1$; $n=-2$, $n'=10$; $h=0.5$, $h'=-2.5$.}
\label{fig:SO10b}
\end{figure}

Figs.\ \ref{fig:SO10a} and \ref{fig:SO10b} refer to the parameters
\bea
&& v=0.5 \ , \ v'=1, \quad \lambda=1,\ \lambda' = 1, \nn \\
&& h=0.5, \ h'=-2.5 \ .
\eea
Fig.\ \ref{fig:SO10a} assumes the winding numbers $n=1$, $n'=-5$,
while Fig.\ \ref{fig:SO10b} addresses the even more exotic case
$n=-2$, $n'=10$ (which is certainly unstable under quantum
fluctuations). This entails larger absolute values of $a(r)$
than in the previous plots, with $a(r \to \infty) = -2$ and $4$,
respectively. We did not encounter any conceptual or numerical
difficulties in dealing with these extraordinary cases.

These scenarios of strong windings of the $\chi$-field confirm
previous observations, now in an amplified form. At small $r$,
the $\chi$-profile function $\xi (r)$ is kept close to zero,
in a range which grows monotonically with $|n'|$; for $|n'|=10$
this range extends up to $r \gtrsim 1$.

Moreover, these plots further confirm that the overshooting effect of
$\phi(r)$ can also occur in the case with U(1)$_{Y'}$ gauge symmetry.
The comparison with Fig.\ \ref{fig:CS5} shows that also this effect is
enhanced by an increasing value of $|n'|$.

Part of the plots in this section include some curves, which do not
fulfill the conditions (\ref{condSSB}), in particular $\mu^{2}$ or 
$\mu'^{\, 2}$ may turn positive, but we do not observe discontinuities
in the set of profile functions when this happens.

\section{Summary and Conclusions}

We have studied an extension of the Standard Model, where the
exact $B-L$ conservation is explained by a U(1)$_{Y'}$ gauge
symmetry. Gauge anomalies are avoided by adding a right-handed
neutrino $\nu_{R}$ to each generation. $\nu_{R}$ obtains an individual
mass term when we add a 1-component additional Higgs field $\chi$,
with quantum number $(B-L)_{\chi} = 2$, and use it to build a
Majorana-Yukawa term $\propto \chi \nu_{R}^{T} \nu_{R}$.
For the standard Higgs field $\Phi$ and for $\chi$ we assume
(renormalizable) quartic potentials --- with negative quadratic
terms that imply spontaneous symmetry breaking --- as well as a
term $- \kappa \Phi^{\dagger} \Phi \chi^{*} \chi$.

We then studied the coupled field equations of $\Phi$ and $\chi$
and the U(1)$_{Y'}$ gauge field ${\cal A}_{\mu}$.
In particular we made an ansatz for cosmic string solutions,
and applied numerical methods to obtain the radial profile
functions with a variety of winding numbers. This analysis
also includes the case where this model is considered part
of the SO(10) GUT, such that the winding numbers of $\Phi$ and $\chi$,
$n$ and $n'$, are related as $n'=-5n$. Since $n,\, n' \in \Z$,
this requires a large value of $|n'|$.

Taking the VEV of $\Phi$ as a reference
to fix the energy scale suggests that the cosmic strings
that we obtain are very thin, typically with radii
$r \in {\cal O}(10^{-3})\, {\rm fm}$. The solutions that we found
do not lead to any objection against the possible
existence of such (hypothetical) cosmic strings.

In some cases we compared the Higgs field profiles in the presence
and absence of the ${\cal A}_{\mu}$ gauge field. The difference,
tends to be modest, but involving ${\cal A}_{\mu}$ further reduces
the radii of the cosmic strings.

Also the $\Phi$-$\chi$ coupling constant $\kappa$ has only a
mild influence on the solutions. Its impact is most manifest
in the regime of about half of the cosmic string radius,
where the profile functions have their maximal slopes,
and it mainly affects the $\Phi$-profile.

Increasing winding numbers, in particular the increase of
$|n'|$, keeps the profile function of $\chi$ close to zero
at very small $r$. This effect was systematically observed:
$n' \neq 0$ implies $\chi (r=0)=0$, and increasing $|n'|$
keeps $| \chi |$ small next to the core of a cosmic string.

Originally we were interested in the possibility of ``co-axial cosmic
strings'', where the profile functions of $\Phi$ or $\chi$ would be
negative in some range inside the cosmic string. This would have been
a novelty in the literature, but an extensive search
did not lead to any solution of this kind.

We did, however, find the opposite behavior: in some cases,
the standard Higgs profile ``overshoots'', {\it i.e.}\ inside
the string it can take a value, which is larger than its
VEV far from the cosmic string.
\ \\

\noindent
    {\bf Acknowledgments:} We are indebted to
Eduardo Peinado, Jo\~{a}o Pinto Barros and
Uwe-Jens Wiese for instructive discussions. We thank the organizers
of the {\it XXXV Reuni\'{o}n Anual de la Divisi\'{o}n
de Part\'{\i}culas y Campos} of the {\it Sociedad Mexicana
de F\'{\i}sica}, where this talk was presented by VMV.
This work was supported by UNAM-DGAPA through PAPIIT project IG100219,
``Exploraci\'{o}n te\'{o}rica y experimental del diagrama de fase de
la cromodin\'{a}mica cu\'{a}ntica'', and by the
Consejo Nacional de Ciencia y Tecnolog\'{\i}a (CONACYT). \\
  
\end{multicols}
\medline
\begin{multicols}{2}

\end{multicols}
\end{document}